\documentclass[review]{elsarticle}

\usepackage{lineno,hyperref}
\usepackage{soul}
\usepackage{color}
\usepackage{graphicx}
\usepackage{graphics}
\usepackage{adjustbox}
\usepackage{ulem}
\usepackage{amsthm}
\usepackage{subfigure}
\modulolinenumbers[5]
\usepackage{blindtext}
\usepackage{amssymb,amsmath}
\usepackage[utf8]{inputenc}

\journal{Journal of \LaTeX \ Templates}

\biboptions{authoryear}

\begin{document}

\begin{frontmatter}

%\title{Mid-latitude Ionospheric Trough using GIM's during Solar Cycle 24}
 
\title{Ionospheric and geomagnetic response to the total solar eclipse on 21 August 2017}



\author[unlp,conicet]{Amalia Meza\corref{mycorrespondingauthor}}
\cortext[mycorrespondingauthor]{Corresponding author}
\ead{ameza@fcaglp.unlp.edu.ar}

\author[ialp,conicet]{Guillermo Bosch}
\author[unlp,conicet]{Mar\'ia Paula Natali}
\author[unlp,trelew]{Bernardo Eylenstein}

\address[unlp]{Laboratorio de Meteorolog\'\i{}a espacial, Atm\'osfera terrestre, Geodesia, Geodin\'amica, dise\~no de Instrumental y Astrometr\'\i{}a (MAGGIA), Facultad de Ciencias Astron\'omicas y Geof\'\i{}sicas (FCAG),
Universidad Nacional de La Plata (UNLP), Paseo del
Bosque s/n, B1900FWA, La Plata, Argentina}

\address[conicet]{Consejo Nacional de Investigaciones Cient\'\i{}ficas y T\'ecnicas (CONICET), Godoy Cruz 2290, C1425FQB, Buenos Aires, Argentina}

\address[ialp]{Instituto de Astrof\'isica de La Plata (UNLP - CONICET), La Plata, Argentina}
\address[trelew]{Observatorio Geofísico Trelew,
Trelew, Chubut, Argentina.}


\begin{abstract}
Solar eclipses provide an excellent opportunity to study the effects of a sudden localized change in photoionization flux in the Earth's ionosphere and its consequent repercussion in the Geomagnetic field. We have focused on a subset of the data available from the North American 2017 eclipse in order to study VTEC measurements from GNSS data and geomagnetic field estimations from INTERMAGNET observatories near the eclipse path. Our simultaneous analysis of both datasets allowed us to quantify the ionosphere and magnetic field reaction to the eclipse event with which allowed us to compare how differently these take place in time. We found that studying the behaviour of VTEC differences with respect to reference values provides better insight of the actual eclipse effect and were able to characterize the dependence of parameters such as time delay of maximum depletion and recovery phase.  We were also able to test models that link the ionospheric variations in a quantitative manner. Total electron content depletion measured from GNSS were fed into an approximation of   Ashour-Chapman model at the locations of geomagnetic observatories and its predictions match the behaviour of magnetic field components in time and magnitude strikingly accurately.
\end{abstract}

\begin{keyword}
\texttt{total solar eclipse, VTEC from GNSS, geomagnetic field variation}
\sep \LaTeX\sep Elsevier \sep template
\MSC[2010] 00-01\sep  99-00
\end{keyword}
\end{frontmatter}

\section{Introduction}

The geomagnetic field exhibits variations in timescales that range from fractions of seconds to millions of years. Some of them are internally generated (in the Earth's core, or in the Earth's crust) and others are external, created in the ionosphere-magnetosphere system. Among these the regular daily variation of geomagnetic field during quiet time periods is a common feature of geomagnetic field measurements; the current system associated with the geomagnetic daily variation is typically termed the solar quiet (Sq) current system. The Sq current system appears due to the ionospheric wind dynamo. 

The ionospheric dynamo is produced by movement of charged particles of the ionosphere across Earth's magnetic field. Tidal effects of the Sun and the Moon and solar heating are both responsible  for this effect. It is therefore controlled by two parameters: the distribution of winds and the distribution of electrical conductivity in the ionosphere. The orbital parameters of Earth, Moon, and Sun; the solar cycle; solar flares; and solar eclipses, are some of the external sources  that influence the ionospheric dynamo. In particular any process that alters ionospheric conductivity affects the electric current.

On the illuminated side of Earth the dominant source of ionization is solar UV radiation. 
Solar flares and eclipses are very important and unique events to analyze the dynamo current answer to short timescale perturbations. Solar flares emit far UV together with soft and hard X-rays that penetrate deeper in the atmosphere, temporarily ionizing the E region and D region.  A solar eclipse produces the opposite effect on ionospheric conductivity, as ionization decreases when the Moon's shadow crosses the Earth's atmosphere. Recombination of ionospheric electrons and ions in the absence of light quickly reduces the conductivity.

Solar eclipses can be accurately predicted, therefore its effects on the ionospheric plasma and geomagnetic field can be planned to be studied using complementary techniques, e.g GNSS, ionospheric radiosonde, magnetometers, etc. A solar eclipse produces an abrupt variation of the atmospheric conditions. During this event the photoionization is strongly reduced, the temperature of the atmosphere also falls and a cold spot with a well-defined edge can then be defined. After that, the photoionizing flux recovers its previous magnitude,  and the atmosphere is heated again to the diurnal level \citep{KnizovaandMosna2011}.

Consequently, solar eclipses produce a series of phenomena that can be studied, such as: temporal and spatial analysis of the sudden electron density (or total electron content)  decay \citep{Chernogor2013, Kumaretal2013,LyashenkoandChernogor2013}, the abrupt geomagnetic variation \citep{KimandChang2018} and the generation of gravity and acoustic waves \citep{JAKOWSKI2008,Chenetal2011}.
The changes in the electron density distribution and the total electron content (TEC) have been studied using different techniques such as the Faraday rotation measurements \citep{Tyagietal1980,DASGUPTA1981}, ionosonde networks \citep{Leetal2009,Kurkin2001}, incoherent scatter radars \citep{MacPhersonetal2000,CherniakandLysenko2013}  and  GNSS systems \citep{Afraimovich1998,  Dingetal2010,Cherniaketal2018}.

Eclipse effects  is considered an external source of the geomagnetic field variation, as detected in the middle of twentieth century \citep[e.g.][]{Katoetal1956}. Later, other studies highlighted the relationship between the geomagnetic component variability and the electric current obstruction effect. One of the first studies was presented by \cite{Takeda1984}, in which they show signatures of additional currents and fields generated by the obstruction of Sq  current system due to the ionospheric conductivity depression during the eclipse. A decrease or increase  on the magnetic field component's have been reported by many authors \citep[e.g.][]{Nevalinna1991,Brennes1993,malin1999}. In spite of all of them taking place during different local time, and at different geographical position, the eclipse effect on the geomagnetic field is very evident. 

A few investigations carried out simultaneous measurements of ionospheric and magnetospheric parameters to study the effect of the solar eclipse on them \citep[e.g.][]{Walker1991} that employ measurements of ionograms,  magnetograms, and microbarographs to study the solar eclipse of March 18th, 1988 in East Asia.  \cite{Momani2010} used GPS, Incoherent Scatter Radar and Earth's magnetic field observations the Earth's magnetic field to study the total solar eclipse on August 1st 2008 over Northern Hemisphere. 

In particular the effect of the total eclipse of the Sun on 21 August 2017  on the terrestrial ionosphere was studied exhaustively by several authors \citep[e.g.][]{Cherniaketal2018,Dang2018, Goncharenko2018, Bullett2018,Reinisch2018,Cnossen2019,Wang2019}. The latter concludes that throughout the course of the moon's shadow over a particular observation site, the disturbance winds at the site change direction and consequently their effects on the electron densities of the F2 region also vary. These winds push the plasma down during the eclipse and transport it up to the upper ionosphere after the eclipse. The combination of chemical processes, wind transport, and ambipolar diffusion cause the time lag and the asymmetric characteristic (rapid decline in N$_e$ and slow recovery from eclipse effects) of the topside ionosphere. Consequently, geographic position, local time, geomagnetic conditions constrain the rate of electron content depletion and its corresponding recovery times.

However, almost all studies mentioned above have studied either ionospheric or geomagnetic effects independently.
\cite{Hvozdara2002} proposed a mathematical model based on the classical Ashour-Chapman model to explain the geomagnetic field components' variation. They quantify these in terms of the position of both the quasi-circular spot of the ionospheric conductivity decrease and the location of the geomagnetic observatory. 
In this work we propose to simultaneously study both  ionospheric and geomagnetic response to the 2017 North America Solar Eclipse, combining information provided by GNSS measurements and quantifying its relation to observed geomagnetic perturbations.

\section{Data and Methodology} \label{sec:data}

\begin{figure}[htbp]
\centering
\includegraphics[ width=0.8\textwidth]{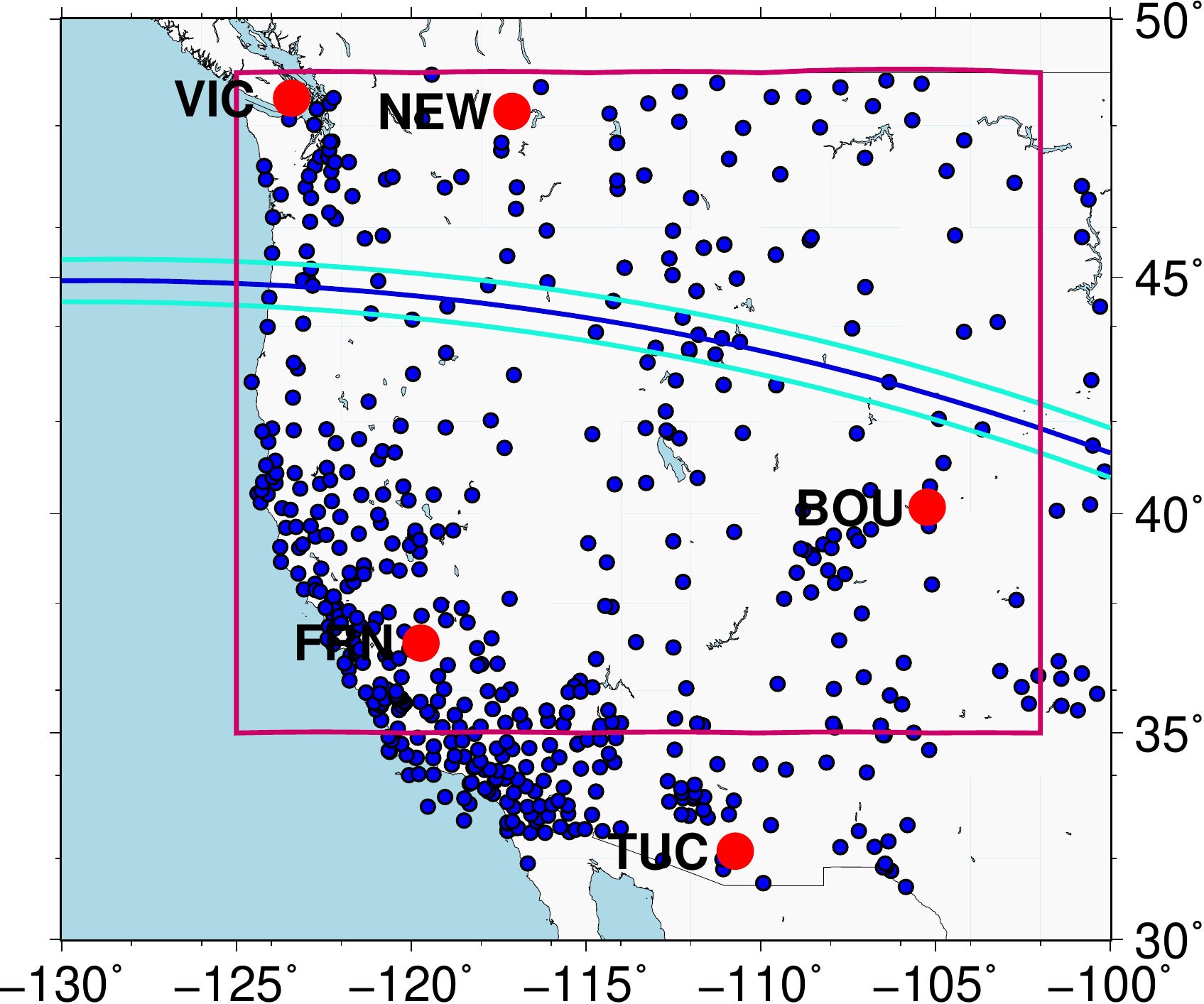}

\caption{GNSS stations (blue circles) and geomagnetic observatories (red circles). The blue and light-blue lines  correspond to the total and the limits of eclipse totality  path, respectively. The red rectangle indicates the area where VTEC maps were calculated}
\label{mapa}
\end{figure}

The VTEC data computed using GNSS measurements and the three geomagnetic field component using magnetometers observations are used in this analysis. A brief description of these data and their variability are presented in this section. Figure \ref{mapa} shows the geographical distribution of the GNSS stations and the location of the geomagnetic observatories. Victoria, Newport, and Boulder geomagnetic observatories lie close to the totality path at a distance shorter than 500 km, so we will restrict our analysis of the North America 2017 eclipse to the area relevant to these observatories indicated with a red rectangle in Figure \ref{mapa}.

\subsection{The VTEC GNSS}
\label{VTEC subsection}

To analyze the response of the ionosphere to the  sudden decrease of radiation from the sun during the solar eclipse, the vertical total electron content (VTEC) is computed. The VTEC values are obtained from the observations recorded by more than 400 ground-based GNSS receivers located in Western United States. All stations belong to the NOAA Continuously Operating Reference Stations (CORS) Network (NCN), managed by NOAA/National Geodetic Survey (ftp://geodesy.noaa.gov/cors/rinex/). 

These observations  were  pre-processed  with  the  Bernese  GNSS Software  version  5.2  \citep{Dach2015}, using models recommended by the International Earth Rotation and Reference Systems Service (IERS) \citep{Petit2010}. Ocean tidal loading  corrections were applied, following \cite{Letellier2004}, together with atmospheric tidal loading displacements provided  by  \cite{vanDamandRay2010},  and  absolute  phase-centre  corrections  for satellites and receivers, as issued by the IGS. 

The Bernese GNSS Software was modified to obtain the phase-code delay ionospheric observable ($\tilde{L}_{I,\mathrm{arc}}$, Equation \ref{eq:Larc}) along with the geographic latitude and the sun-fixed longitude of the ionospheric pierce point, zenith distance ($z'$), azimuth angle, and time for each satellite over each GNSS station displayed in Figure 1.

The ionosphere is approximated by a single shell of infinitesimal thickness with equivalent STEC, located at 450 km above the Earth surface. The intersection point of the line receiver-satellite with the ionospheric layer is  named ionospheric pierce point.  An obliquity factor ($1/\cos z'$ ) is used to map VTEC into STEC (integrated electron density along the signal path); being $z'$ the zenithal distance of the slant path at the ionospheric piercing point.

\begin{equation}
\mathrm{STEC}=\frac{1}{\cos z'} \mathrm{VTEC}
\label{eq:stec}
\end{equation}

The code-delay ionospheric observable is modeled using an arc-dependent bias, $ \tilde{c}_{\mathrm{arc}}$, which accounts for receiver and satellite inter-frequency bias and the ambiguity term. Following \cite{Ciraolo2007} and \cite{Meza2009} the observation equation can be written as:
\begin{equation}
\tilde{L}_{I,\mathrm{arc}}=\mathrm{STEC}+\tilde{c}_{\mathrm{arc}}+\varepsilon _{L}
\label{eq:Larc}
\end{equation}
where $\tilde{L}_{I,\mathrm{arc}}$ is in  TECU (10$^{16}\, el$ m$^{-2}$).
Daily solutions are computed to estimate the arc-dependent bias which is therefore removed from $\tilde{L}_{I,\mathrm{arc}}$ to obtain STEC and VTEC through equations \ref{eq:Larc} and \ref{eq:stec} respectively.

\begin{figure}[htbp]
\centering
\includegraphics[ width=\textwidth]{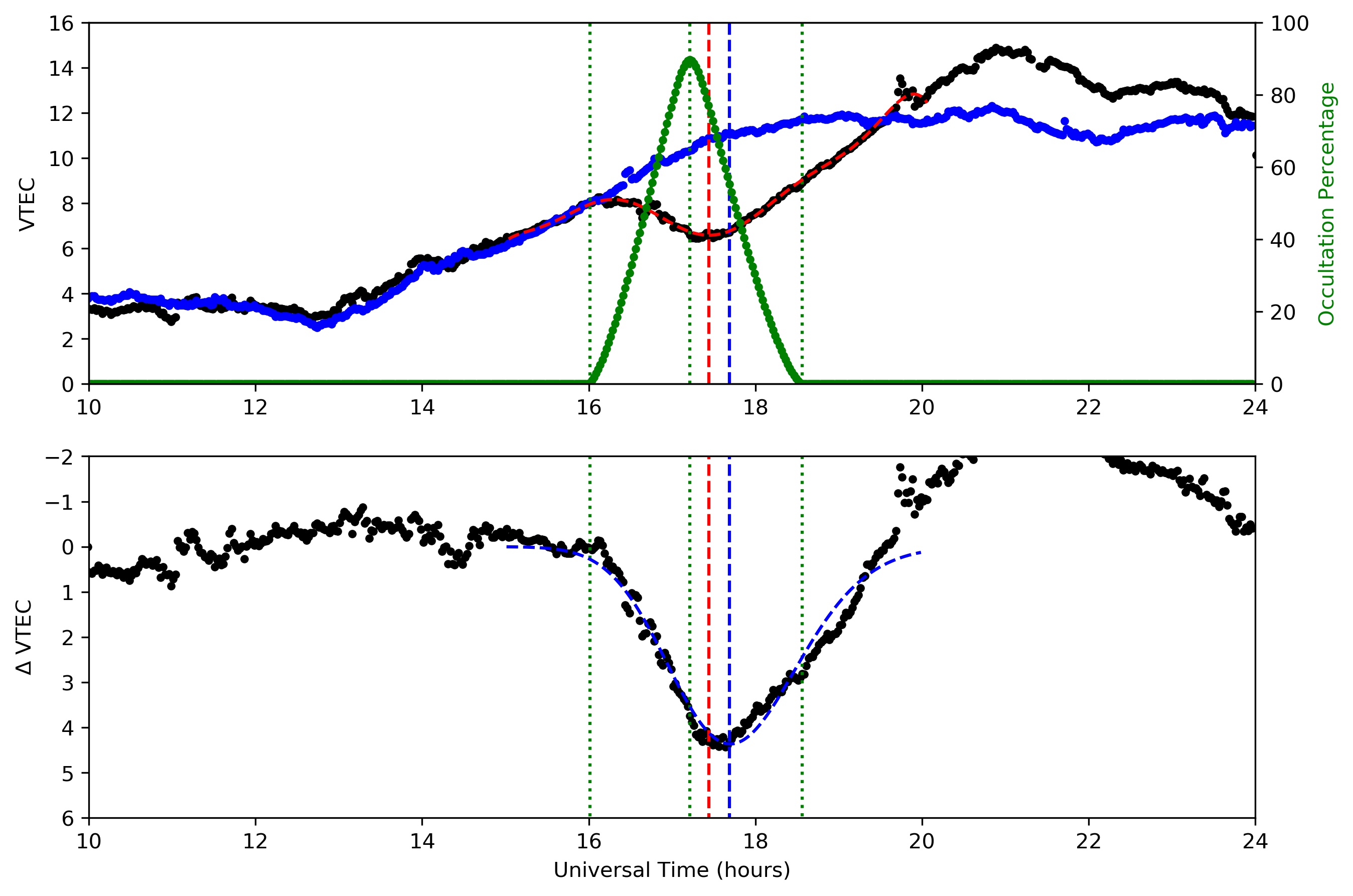}
\caption{Top panel: Black dots show VTEC values together with polynomial fit (red dashed curve) to the VTEC variations during eclipse and the red dotted line indicates the time of the maximum VTEC drop ($t_1$). Blue dots indicate the masked average VTEC values from previous and following days as described in text. Green curve shows the solar obscuration during the eclipse (referenced at right axis), with green dashed vertical lines indicating the first contact, maximum occultation and last contact. Bottom panel: To ease comparison with top panel, the y-axis has been inverted as $\Delta$VTEC is defined positive. Black dots show $\Delta$VTEC values calculated from the difference between reference and eclipse days. The best fit using an exponentially modified Gaussian distribution is plotted as a blue dashed line. The blue dashed vertical line indicates the time of the maximum $\Delta$VTEC derived from the fit ($t_2$). }
\label{Fig:Analysis}
\end{figure}

The derived VTEC determinations were analyzed with an ad-hoc Python code that iterates selecting data relevant to individual one degree longitude and latitude bins. Within each bin the code performs the following tasks:
\begin{enumerate}
    \item Calculates the time evolution of the VTEC values averaging along 1.5 minute intervals all VTEC determinations for the corresponding coordinate pair. (VTEC$_{\mathrm{Ecl}}$, black points in Figure \ref{Fig:Analysis} upper panel).
    \item Uses astronomical ephemeris (PyEphem, https://rhodesmill.org/pyephem/) to derive timing and percentage of eclipse obscuration at a reference altitude of   120 km \citep{Yamazaki2017} (green dotted vertical lines and green points respectively in Figure \ref{Fig:Analysis} upper panel). This reference height was chosen to match the altitude of the ionosphere E-layer, where the electric field and Sq currents are mostly affected by changes in electron density during the eclipse. The timing of maximum occultation ($t_0$) will be used as reference for estimating time delays of other ionospheric effects.
    \item Fits a polynomial function to the VTEC variations (red dashed curve) during the eclipse to determine the timing of the maximum drop measured on the VTEC curve due to the occultation ($t_1$, red dotted vertical line in Figure \ref{Fig:Analysis}) and its corresponding delay with respect to $t_0$ (${\Delta t}_1=t_1-t_0 $).
    \item Calculates a masked average of VTEC values from the immediately previous and following days (VTEC$_\mathrm{Ref}$, blue points in Figure \ref{Fig:Analysis} upper panel). The masked average consists of masking time intervals without data and consequently using the only available data point when the other one is missing or performing the average if both are present. In this way we can obtained a well sampled VTEC reference to derive the change in VTEC ($\Delta$VTEC, black points in Figure \ref{Fig:Analysis} bottom panel) by comparing the observed behaviour against VTEC$_{\mathrm{Ecl}}$.
    \item Fits a Skewed Gaussian distribution (SkG,  \cite{ASHOUR2010341} and references therein) to the $\Delta$VTEC values during the eclipse event (blue dashed line in Figure \ref{Fig:Analysis} bottom panel) in order to derive both the timing ($t_2$, blue dotted vertical lines) and the maximum value of the VTEC difference, defined as $\Delta$VTEC$_{\mathrm{max}}$ = max( VTEC$_{\mathrm{Ref}}$ - VTEC$_{\mathrm{Ecl}}$). The use of the SkG allows to account for the varying behaviour of $\Delta$VTEC curve and derive additional parameters such as skewness, which will address the relation between depletion and recovery times.
    \item Obtains spatially resolved information of $\Delta$VTEC$_{\mathrm{max}}$ and ${\Delta t}_2 = t_2 - t_0$, i.e. the time delay between the maximum Solar obscuration and $\Delta$VTEC$_{\mathrm{max}}$.
\end{enumerate}

\subsection{The geomagnetic field }

As stated before, we focused on geomagnetic values corresponding to data collected at the three ground based observatories closest to the totality path. The Boulder (BOU) and Newport (NEW) Observatories are operated by the United States Geological Survey (USGS), and the Victoria (VIC) Observatory is in turn supported by the Geological Survey of Canada (GSC). All three participate in the International Real-time Magnetic Observatory Network (INTERMAGNET). Consequently the geomagnetic values were obtained from the INTERMAGNET data site. Considering that the eclipse signature in the geomagnetic field, as seen by \cite{malin1999} and  \cite{Hvozdara2002}, is noticed as a smooth variation of its components within a 1-hour interval, we chose to work with the available data at a sampling rate of 1 minute. These values are usually obtained by applying a digital Gaussian filter to a higher sample rate data set centered in the minute, thus eliminating short-term disturbances such as errors due to the instrument. The retrieved data were already originally available in the three geomagnetic field components of interest ($X, Y, Z$). In order to reject superposed disturbances of much shorter period, the minute data will be further smoothed by a 15-min window moving average.

The geomagnetic field variability is defined as the difference between the values obtained during the eclipse event and the reference values.  The latter are obtained calculating the mean value of the five nearest geomagnetic quiet days \citep{Momani2010}. 

To examine patterns of the geomagnetic field variations induced by solar eclipse, we analyzed data within a two hours time interval centered in the maximum occultation .  In order to eliminate the intrinsic regular daily variabilities of the eclipse  day and the reference day, during the two hour window selected,  the linear trend was removed using a first order polynomial fit \citep{malin1999}. 
Finally the geomagnetic field variations, produced by the solar eclipse,  $\Delta X$, $\Delta Y$ and $\Delta Z$ are computed and shown in the left column of Figure \ref{variacion geomagnetica}.

\subsection{Relationship between VTEC and geomagnetic field}
\label{Sec:GeomagneticModels}

The classical Ashour-Chapman model, with \cite{Hvozdara2002} modifications,  is considered to analyse the geomagnetic components variability and its relationship with the VTEC variation in the region of eclipse obscuration.

Low-conductivity ionospheric spot is used as the Ashour-Chapman model of a thin current sheet model with the arbitrarily directed undisturbed electric field $\bf{E_{0}}$. In the present work the angle between x-axis (to geographic North) and $\bf{E_{0}}$, $\epsilon$, is different  from zero  and the direction of the equivalent Sq current system is assumed similar to $\bf{E_{0}}$ (in this first approximation the Hall conductivity is not taken into account for  determination). Dedicated Ionospheric Field Inversion (DIFI-3) model, time-varying spherical harmonic representation of the quiet-time Sq and equatorial electrojet field, is used to $\epsilon$ determination (https://geomag.colorado.edu/difi-calculator).

Our analysis is based on the mathematical explanation described in  \cite{Hvozdara2002}, Appendix A. 
In their paper, the authors model the geomagnetic effect due to changes in the local ionospheric conductivity linked to the TEC decrement originated by the eclipse. They do this by means of a cilindrical coordinate system ($r$, $\phi$, $z$) with origin on the eclipse-induced conductivity spot and its z axis normal to the Earth's surface. In this system, the magnetic potential field can be written as:
\begin{equation}
\mathrm{{\Omega }=-I \, a \, \sin \, \phi \, W(r,z)}
\end{equation}
where:
\begin{equation}
\setlength{\jot}{12pt} % affecting the line spacing in the environment
\begin{split}
I &=  \frac{1-\kappa }{1+\kappa }\, I_{0} [A/m] \\
W(r,z) &=\int_{0}^{\infty}s^{-1}J_{1}(sa)J_{1}(sr)e^{-sz} ds
\end{split}
\end{equation}
$J_{1}$ is the Bessel function of the first kind and index 1.

Being $\mathbf{H}$ the disturbing magnetic field which is related with the corresponding potential $\mathbf{\Omega }$ \citep{Ashourandchapman}. Then
\begin{equation}
\mathbf{H}=-\mathrm{grad} \,{\Omega}
\end{equation}
The geomagnetic disturbance is defined by $\mathrm{\mathbf{b}=\mu _{0} \, \mathbf{H}}$ where $\mu _{0}= 400 \, \pi$ and $\mathbf{b}$ is in nT unit. Its cartesian components are:
\begin{equation}
\begin{split}
b_{x}&=\mu _{0} (H_{r} \,\cos \alpha-H_{\varphi} \,\sin \alpha )\\
b_{y}&=\mu _{0} (H_{r} \,\sin \alpha+H_{\varphi} \,\cos \alpha )\\
b_{z}&=\mu _{0} H_{z}
\end{split}
\end{equation}
Table \ref{Table 1} shows the values defined to calculate the geomagnetic disturbance for the different geomagnetic stations. The angle $\epsilon$, the distance from the observatories and the center of the eclipse-induced conductivity spot, $\delta$, and the degree of the TEC decrease caused by the solar eclipse, $\kappa$.

\begin{table}
\centering
\caption{Parameters used in the theoretical model of geomagnetic eclipse-disturbance: $\epsilon$ is the angle between x-axis (to geographic North) and $\bf{E_{0}}$; $\delta$ is the distance from the observatories and the center of the eclipse-induced conductivity spot,  and $\kappa$ is the degree of the VTEC decrement caused by the solar eclipse}
\label{Table 1}
\begin{tabular}{lrrr}
\hline
    &  $\epsilon$   &  $\delta$    & $\kappa$     \\
\hline
VIC & 130 & 390  & 0.63 \\ 
NEW & 116 & 450  & 0.61 \\
BOU & 97  & -206 & 0.57 \\
\hline
\end{tabular}
\end{table}

The magnetic field variations $\mathbf{b}$ $=(b_{x}, b_{y}, b_{z})$ are adjusted for the electromagnetic induction effect, in order to  compare with the geomagnetic field components variability from the geomagnetic observatories \cite{Hvozdara2002}. Consequently, the  $\Delta X$ , $\Delta Y$ and $\Delta Z$, predicted by the model are expected to be 1.5 $b_{x}$, 1.5 $b_{y}$ and 0.3 $b_{z}$ respectively.

\section{Results and  Discussions}
\label{sec:Results}

\subsection{VTEC variation}
\label{subsec:31}

\begin{figure}[htbp]
\centering
\includegraphics[width=\textwidth]{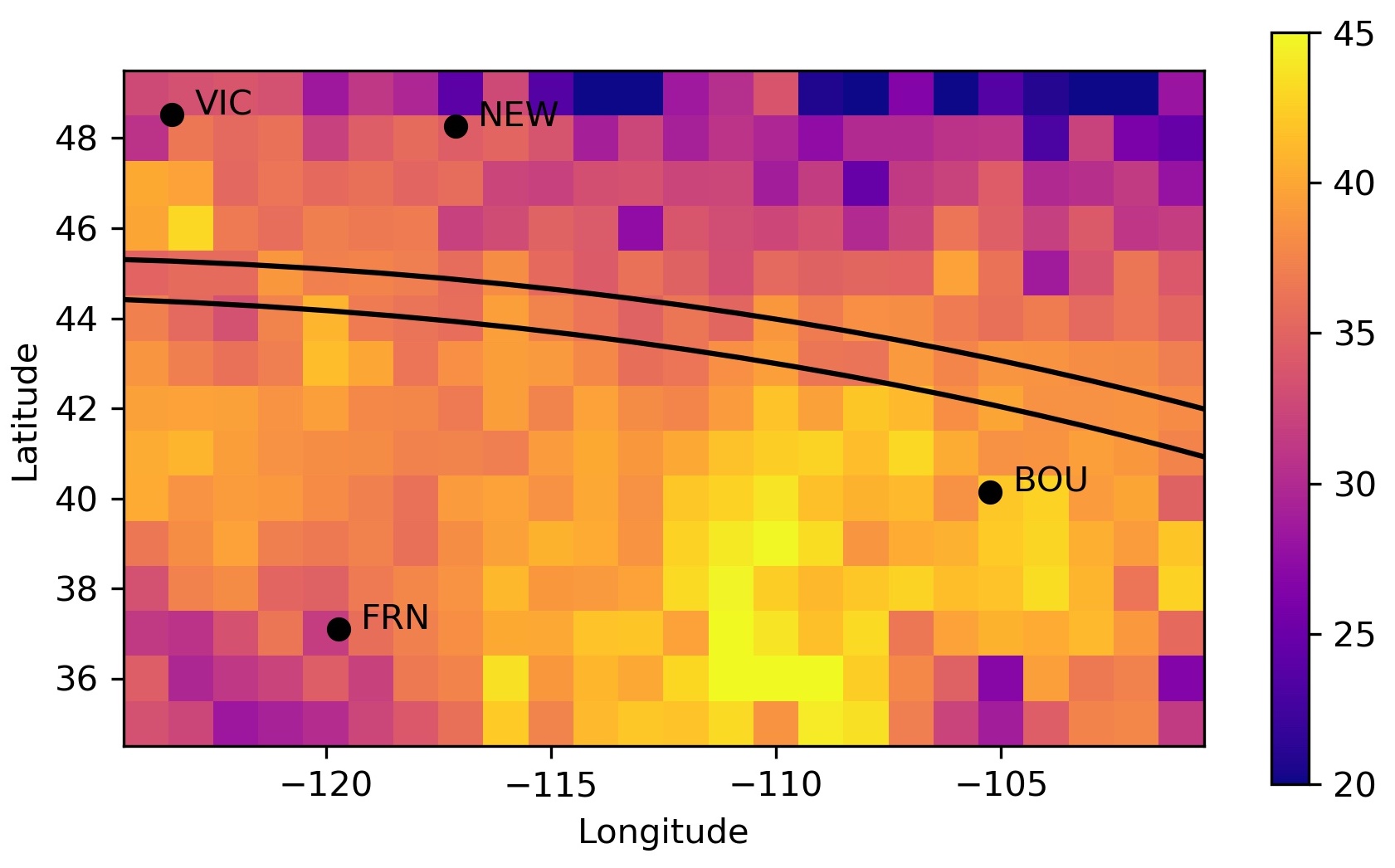}
\caption{Geographical distribution of relative $\Delta$VTEC$_{\mathrm{max}}$ (in percentage units). Lower and upper limits of eclipse totality path, together with locations of geomagnetic observatories are also plotted as reference.}
\label{MDTECU}
\end{figure}

\begin{figure}[htbp]
\centering
\includegraphics[width=\textwidth]{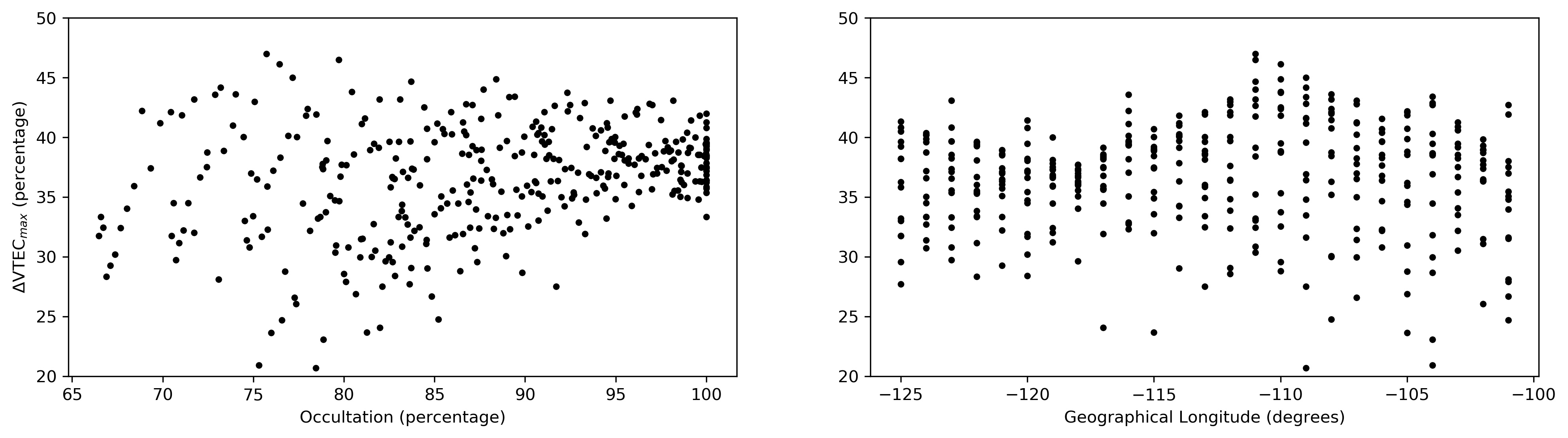}
\caption{Relative $\Delta$VTEC$_{\mathrm{max}}$ plotted against maximum eclipse occultation (left panel) and geographical longitude (right panel).}
\label{regresion}
\end{figure}

Figure \ref{MDTECU} shows the $\Delta$VTEC$_{\mathrm{max}}$ during the eclipse, displayed as a percentage value from the reference VTEC value. No evident trend can be seen longitudinally, besides a local peak about 110\textdegree W. Figure \ref{regresion} shows the $\Delta$VTEC$_{\mathrm{max}}$ as function of the geographical longitude, confirming that no dependence on occultation percentage is present either. \cite{Liu19} studied the effects of an annular eclipse centered in Taiwan and found a strong  correlation between maximum obscuration and $\Delta$VTEC$_{\mathrm{max}}$ in percentage. Their figure 6i shows that $\Delta$VTEC$_{\mathrm{max}}$ is directly proportional to maximum obscuration (ranging between 40\% and 90\%), although there is evidence of saturation at high occultation. Bottom/right panel in Figure \ref{regresion} confirms this, as no visible trend with $\Delta$VTEC$_{\mathrm{max}}$ is present when focusing on occultations larger than 75\% .

There is a noticeable difference among $\Delta$VTEC$_{\mathrm{max}}$ north and south of the totality path. We can interpret this considering that the integral of the electron content in the ionosphere results from the balance between transport, ionization and loss processes; the eclipse reduces the electron temperature, decreases the pressure and consequently induces a downward drift of plasma from the topside ionosphere \citep{Dingetal2010,Cherniaketal2018}. An explanation of the latitudinal distribution of $\Delta$VTEC$_{\mathrm{max}}$ is that the eclipse switches off the ionization source, therefore the recombination is more effective specially at lower latitudes where the neutral mass density is higher \citep{Cherniaketal2018}. 

\begin{figure}[htbp]
\centering
\includegraphics[ width=\textwidth]{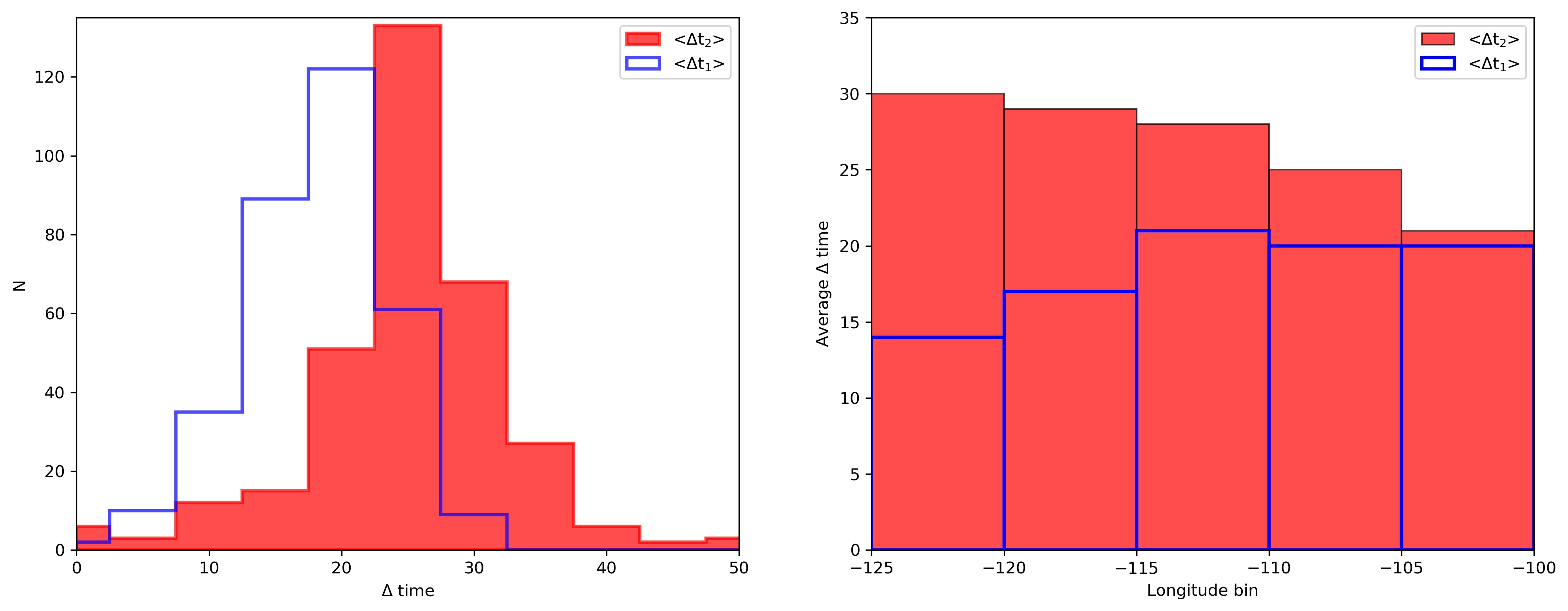}
\caption{Analysis of time delay distribution between maximum eclipse occultation and ionospheric response. Both panels show trends of ${\Delta t}_1$ in blue outline and ${\Delta t}_2$ in filled red.  Left panel highlights the overall difference between time delays and right panel displays distinct dependence of time delays with geographical longitude.}
\label{Fig:DeltaTime}
\end{figure}

Another interesting result worth discussing is the analysis of the time delay between solar occultation and the ionospheric response measured from VTEC. 
The presence of a time delay between occultation and ionospheric response has already been presented in previous researches \citep[e.g.][]{JAKOWSKI2008,Boitman99, Liu19,Cherniaketal2018,Momani2010}; although showing shorter delays. As  outlined in Section \ref{VTEC subsection} we measure time delays from the $\Delta$VTEC curve as this is more strongly linked to the eclipse effect itself. 

The variations in the local VTEC minimum, cited in references above are actually due to a combination of the eclipse occultation and the rapid increase in VTEC expected at mid-morning. The apparent increase in ${\Delta t}_1$ seen in Figure \ref{Fig:DeltaTime} is not due to the eclipse itself but to the fact that the daily VTEC increase in the ionosphere is slowing down as it reaches its peak at about 20hrs UT.
For comparison purposes we also show in Figure \ref{Fig:DeltaTime} the time delay distribution derived from the VTEC curve which resemble values cited in the literature. Furthermore, we were able to reproduce to longitude bin average of the time delay as calculated from the VTEC curve in the right panel of Figure \ref{Fig:DeltaTime}. Blue outline bins show almost identical behaviour as in \cite{Cherniaketal2018}, a small (2 to 3 minutes) offset is present, mostly due to the fact that the cited authors refer to eclipse ephemeris at ground level while we do it 120 km above. Filled red bins show, besides larger values as already mentioned a slow but constant decreasing pattern as the eclipse advances eastward.

\begin{figure}[htbp]
\centering
\includegraphics[width=\textwidth]{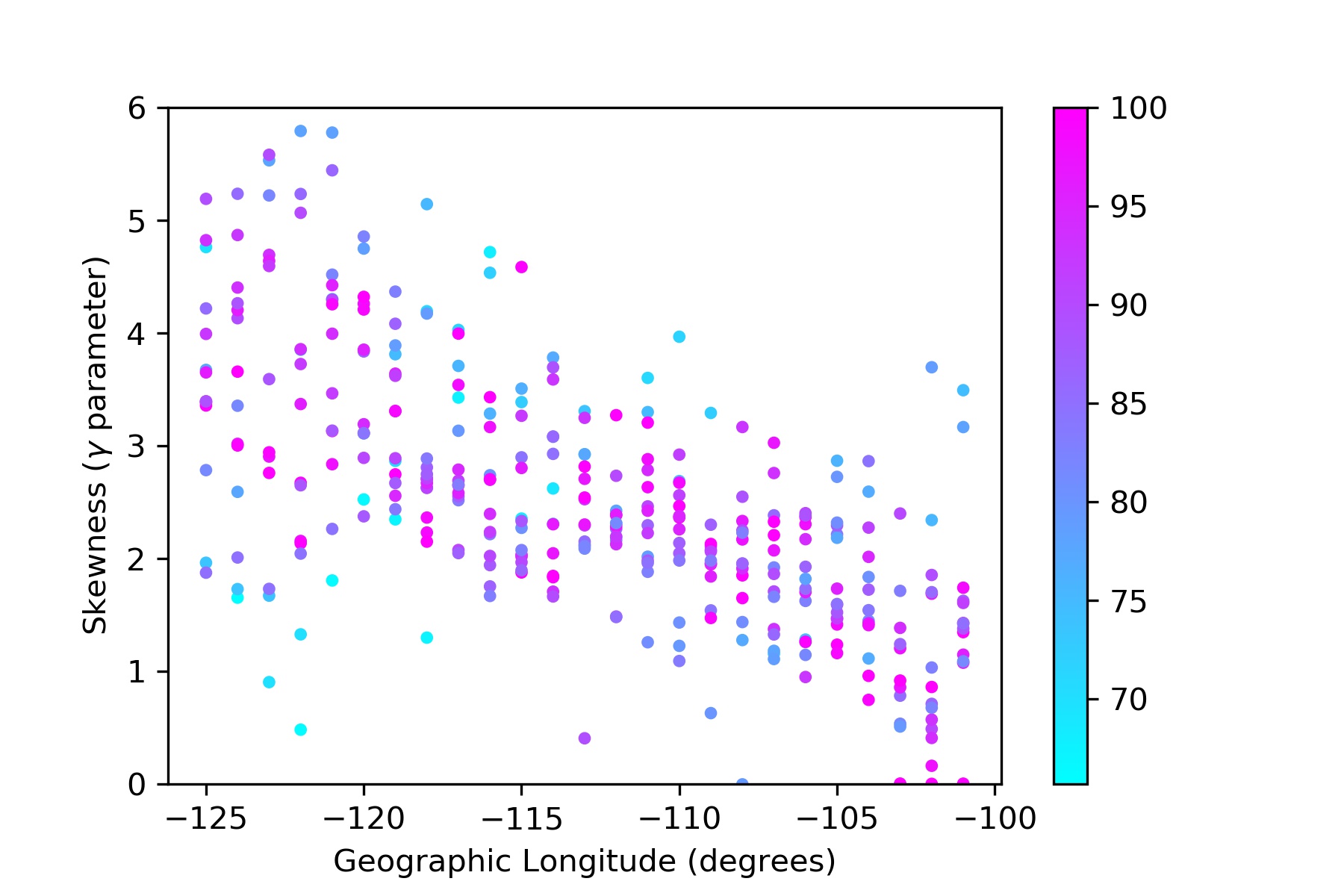}
\caption{Skewness variation plotted against geographic longitude where the strong reduction of asymmetry in the $\Delta$VTEC profile can be seen. We have added a color scale for eclipse occultation making it evident there is no dependence on this parameter.}
\label{skewness}
\end{figure}

We also analyzed the presence of trends among the asymmetry in the $\Delta$VTEC curve measured via the $\gamma$ parameter of the skewed Gaussian profile. A $\gamma$ value of 0 indicates a symmetric Gaussian profile and positive values show recovery times larger than depletion ones.
Figure \ref{skewness} displays a visible trend with $\gamma$ values diminishing asymmetry eastwards even within the scatter. This indicates that the recovery time is noticeable larger than depletion time at the west coast but this difference as the shadow moves eastward. Several effects can contribute to this behaviour, as the moon shadow changes its shape and reduces its apparent speed when its cone axis passes closest to Earth's center.   

\subsection{Geomagnetic field variation}
\label{subsec:32}

The geomagnetic variability during total solar eclipse is analysed over the observatories affected by an occultation larger than 80 \%: VIC has the 89,1 \% at 17:20 UT (10:20 LT), NEW has the 84,6 \% at 17:26 UT (10:26LT), and BOU has the 92,4\% at 17:46 UT (11:46 LT), its geographical locations are shown in Figure 1.

\begin{figure}[htbp]
\centering
\includegraphics[width=\textwidth]{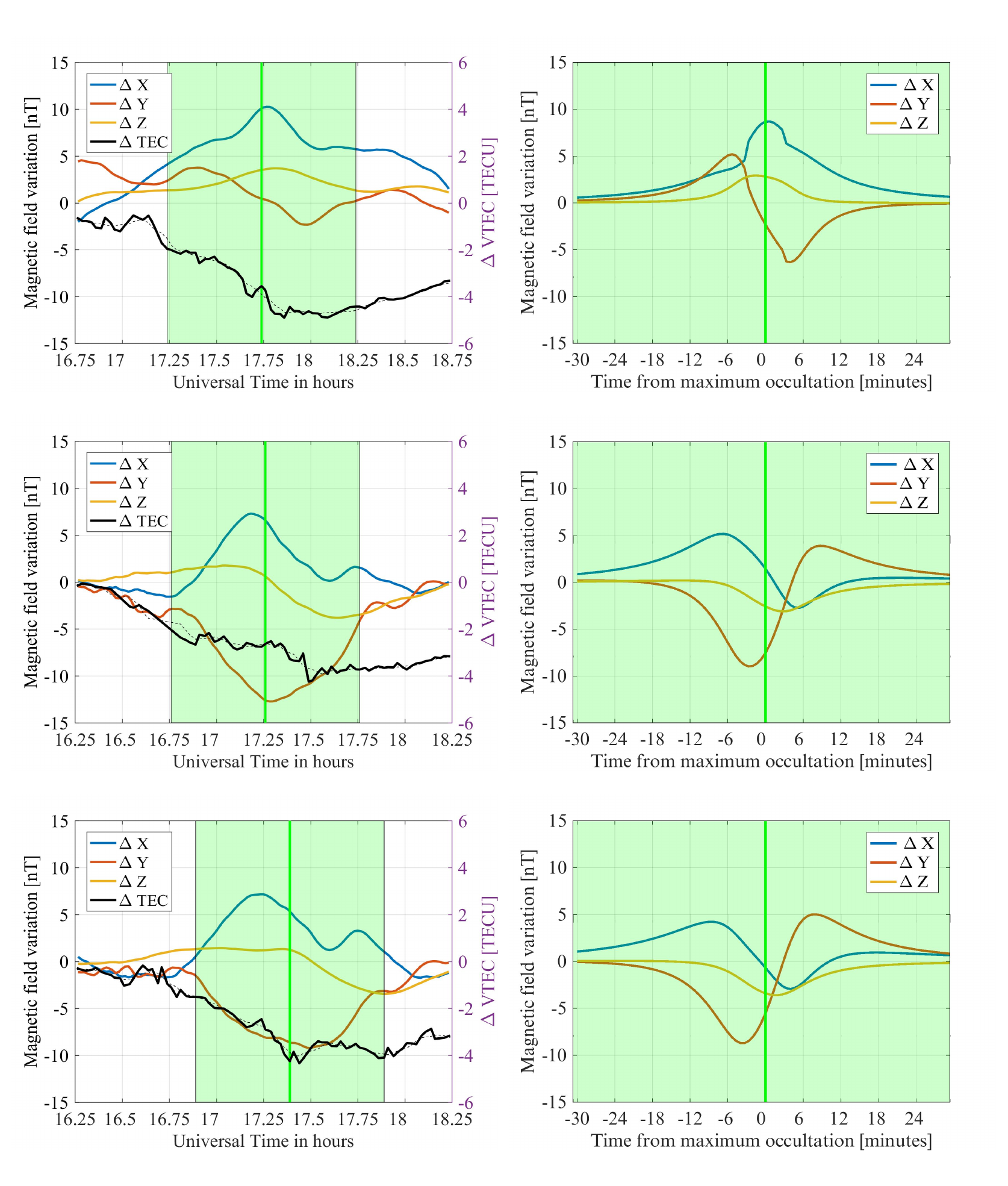}
\caption{Geomagnetic variability detected from BOU, VIC, and NEW observatories. The left column shows the measurements of  $\Delta X$, $\Delta Y$, and $\Delta Z$, in blue, red, and yellow color respectively. The black curve shows the $\Delta$VTEC$_{\mathrm{max}}$ measurement. The right column shows the geomagnetic variability based on the Ashour-Chapman thin current sheet model.}
\label{variacion geomagnetica}
\end{figure}

In this section we will discuss the comparison of the observed geomagnetic variability and the one proposed by the theoretical model described in Section \ref{Sec:GeomagneticModels} and displayed in the right column of Figure \ref{variacion geomagnetica}.
These modeled geomagnetic disturbances along the Cartesian components are defined along one hour interval centered at the time of maximum occultation ($t_0$), which is highlighted by the green box .

Top row in Figure \ref{variacion geomagnetica} displays BOU observatory geomagnetic variability:  the $\Delta X$ component shows positive values with its maximum arising near $t_0$, the $\Delta Y$ component has positive and then negative values before and after $t_0$ respectively, and the $\Delta Z$ component has the lower amplitude, showing positive values with its maximum near $t_0$.  Its corresponding right panel shows the very good agreement for this eclipse configuration. 
Middle and bottom  rows (VIC  and  NEW  observatories)  show  similar  behaviours among themselves,  which is  quite  expected  due  to  close  geographical  coordinates and relative location to the eclipse path,  but quite different as the one shown at BOU.  The $\Delta X$ component is positive and its maximum value takes place before $t_0$,  the $\Delta Y$ component is negative and its minimum value happens close to $t_0$, and the $\Delta Z$ Component records lower magnitude variations, where its maximum difference is negative and occurs after $t_0$.  This distinct observed behaviour is still well described qualitatively by the model in their corresponding right panels, but the agreement is not as good as for BOU observatory.

 \cite{malin1999} and \cite{Momani2010} had also found perceptible differences in geomagnetic response at different observatories locations. Regarding the model predictions, dissimilar performance among observatories can be attributed to a combination of observing conditions and model limitations. VIC and NEW stations lie farther from eclipse path and smaller geomagnetic disturbances are therefore expected which might be subject to contamination by other sources of variations. Model assumptions (such as cylindrical symmetry and ionospheric isotropy) can also contribute to these discrepancies.

As a reference, we have also added the $\Delta$VTEC values in the left column of Figure \ref{variacion geomagnetica} which allows a straightforward comparison between the VTEC and geomagnetic variations.

 From the visual inspection of Figure  \ref{variacion geomagnetica} it can be readily seen that eclipse effects impact on the VTEC and the geomagnetic field with very different timescales. We interpret this as due to the fact that the variations of electric current which flow mainly in the E layer are the ones responsible for the slight changes in the geomagnetic field observed from ground level observatories. This implies that the time variation of eclipse effects on the magnetic field depends mostly on processes taking place in the E layer. On the other hand, even though the VTEC determinations are integral determinations over the line of sight, the F2 layer provides the major contribution to it. Therefore, an explanation of the observed delay of the VTEC variability on Figure \ref{variacion geomagnetica} is that in the F2 layer the recombination process time is longer than E layer, because the eclipse effect is controlled by the transport of plasma \citep{Bienstock1970}.

\section{Concluding remarks}  

 This work examines the ionospheric and geomagnetic response to the 2017 North America Solar Eclipse.  The VTEC obtained from GNSS observations were used to compute the VTEC variability during the eclipse; the $\Delta$VTEC, which is the difference between VTEC during the eclipse and reference days, and its maximum ($\Delta$VTEC$_{\mathrm{max}}$) and also the maximum drop of VTEC curve due to the occultation.
The values of $\Delta$VTEC$_{\mathrm{max}}$ range from 20\% to 45\% along the eclipse path within an area of 70\% obscuration, and the maximum values stretch equatorward from the totality.
The $\Delta$VTEC$_{\mathrm{max}}$ lags 22-32 min after the maximal eclipse time (${\Delta t}_1$).
The values of VTEC drop had a 14-22 min time delay after the maximum occultation (${\Delta t}_2$). 

The ${\Delta t}_2$ values and the recovery times concerning the depletion times ( $\gamma$ parameter) are strongly related to local time. They have larger values at westwards longitudes when the eclipse is at mid-morning. When the eclipse is carried out close to midday, the depletion and recovery times, and ${\Delta t}_1$ and ${\Delta t}_2$, become more similar.

For the first time, a total solar eclipse was simultaneously studied from both ionospheric and geomagnetic points of view. The degree of the TEC decrease caused by the solar eclipse was used in a mathematical model based on the Ashour-Chapman model to predict the geomagnetic disturbance. Quantitatively the model and the Cartesian geomagnetic components variabilities of the three geomagnetic observatories, with larger occultation than 84\%, were comparable and consistent.

\section{Acknowledgments}
 We wish to acknowledge the thorough review and useful comments provided by the anonymous referees, which have helped to improve the final version of this article. The results presented in this paper rely on data collected at magnetic observatories. We thank the national institutes that support them and INTERMAGNET for promoting high standards of magnetic observatory practice (www.intermagnet.org).

\bibliography{eclipse_ref}

\end{document}